\documentclass[review, onecolumn,english,american,prl,manuscript,superscriptaddress]{revtex4}
\usepackage[LGR,T1]{fontenc}
\usepackage[latin9]{inputenc}
\usepackage{color}
\usepackage{multirow}
 
\setcitestyle{super}
\usepackage{amsmath}
\usepackage{amssymb}
\usepackage{graphicx}
\usepackage{textcomp}

\usepackage{array}
\newcolumntype{L}{>{\centering\arraybackslash}m{2cm}}

\makeatletter

\ProvideTextCommand{\~}{LGR}[1]{\char126#1}

\@ifundefined{textcolor}{}
{%
 \definecolor{BLACK}{gray}{0}
 \definecolor{WHITE}{gray}{1}
 \definecolor{RED}{rgb}{1,0,0}
 \definecolor{GREEN}{rgb}{0,1,0}
 \definecolor{BLUE}{rgb}{0,0,1}
 \definecolor{CYAN}{cmyk}{1,0,0,0}
 \definecolor{MAGENTA}{cmyk}{0,1,0,0}
 \definecolor{YELLOW}{cmyk}{0,0,1,0}
}

\usepackage[english]{babel}

\def\url#1{}
\addto\captionsenglish{%
}

\usepackage[none]{hyphenat}
\usepackage{multirow}

\usepackage{babel}

\makeatother

\usepackage{babel}

\begin{document}

\title{2D Magnetic Heterostructures: Spintronics and Quantum Future}
\author{Bingyu Zhang}
\affiliation{Department of Electrical and Computer Engineering, University of
Florida, Gainesville, Florida 32611,
USA.}
\author{Pengcheng Lu}
\affiliation{Department of Materials Science and Engineering, University of
Florida, Gainesville, Florida 32611,
USA.}
\author{Roozbeh Tabrizian}
\affiliation{Department of Electrical and Computer Engineering, University of
Florida, Gainesville, Florida 32611,
USA.}
\author{Philip X.-L. Feng}
\author{Yingying Wu}
\thanks{Correspond to: yingyingwu@ufl.edu}
\affiliation{Department of Electrical and Computer Engineering, University of
Florida, Gainesville, Florida 32611,
USA.}



\begin{abstract}
The discovery of two-dimensional (2D) magnetism within atomically thin structures derived from layered crystals has opened up a new realm for exploring magnetic heterostructures.\,This emerging field provides a foundational platform for investigating unique physical properties and exquisite phenomena at the nanometer and molecular/atomic scales. By engineering 2D interfaces using physical methods and selecting interlayer interactions, we unlock the potential for extraordinary exchange dynamics. This potential extends to high-performance and high-density magnetic memory applications, as well as future advancements in neuromorphic and quantum computing. This review delves into recent advances in 2D magnets, elucidates the mechanisms behind 2D interfaces, and highlights the development of 2D devices for spintronics and quantum information. Particular focus is placed on 2D magnetic heterostructures with topological properties, promising for a resilient and low-error information system.\,Finally, we discuss the trends of 2D heterostructures for future electronics, considering the challenges and opportunities from physics, material synthesis, and technological prospective. 
\end{abstract}
\maketitle

\newpage

\section{Introduction}
The advent of 2D van der Waals (vdW) materials\cite{wu1BP,wu2BP,wu3SC,wu4Ising} has sparked a revolution in the field of magnetism, primarily due to the substantial influence of the dimensionality confinement and size effects upon spin textures and dynamics, e.g., quantum fluctuations. These 2D layered magnetic materials (2D magnets) have garnered immense attention, driven by both their fundamental significance and facile integration into multi-layer heterostructures. Ever since the experimental discovery of 2D magnets with intrinsic magnetism in 2017, like CrI$_3$ and Cr$_2$Ge$_2$Te$_6$ \cite{gong2017discovery,huang2017layer}, 2D magnets family has expanded significantly over the years. Unlike ultrathin film systems grown on a substrate (e.g., CoFeB/MgO), these 2D magnets possess a naturally layered structures, high crystallinity, and weak coupling to any transferred substrate. Furthermore, the reduced coordination effects\cite{wang2020prospects}, e.g., finite size effect, seen in ultrathin films are less pronounced in these layered magnets, thanks to their weak interlayer interactions. Consequently, these 2D magnets serve as tailored testbeds for exploring the pure dimensional transition of magnetism from 3D to 2D. Additionally, these magnets exhibit high responsivities to external stimuli, such as gate voltage, molecule adsorption, and neighboring materials. The unique controllability of magnetism by electric ways opens up possibilities for creating heterostructures and devices for applications in spintronics and memory technology.

However, it is worth noting that the magnetic properties of individual 2D magnets are typically monotonous and imperfect, which poses significant limitations for further applications. Therefore, an effective strategy involves modulation of their magnetism through heterostructure construction, which can tune the electronic band structures of 2D magnets via interfacial interactions. Artificial structures can be assembled to manipulate and enhance their magnetism through magnetic coupling. Furthermore, these 2D magnets can be harnessed to induce valley polarization and spin splitting in many nonmagnetic 2D materials, such as monolayer graphene\cite{wu2020large} and topological insulators\cite{pan2022efficient, yang2020termination,wang2020topological}.

Based on these magnetic heterostructures, 2D spintronics is an emerging field to realize devices such as magnetic tunneling junctions (MTJ), spin field-effect transistors and memoristors\cite{2Dapplication1,2Dapplication2,2Dapplication3,2Dapplication4}. 2D spintronics has unwrapped innovation and compelling opportunities which not only improve their performance but also diversify the functionality of electronic devices. For example, imprinted magnetic skyrmions can only exist at 2D interfaces\cite{wu2022van}, which can be used as parallel information storage channels. When combined with electrical control, switch of skyrmion on/off states can be realized\cite{skyrmiononoff1,skyrmiononoff2}.\,Besides, 2D magnetic heterostructures for unconventional computing is an emerging field.\,One examples is for neuromorphic computing, which traditionally employs 3D bulk materials or nonlayered thin films, and thus the resulting device size is either difficult to scale down for high density integration or suffering from lattice mismatch problems.\,The emergence of 2D magnets offers a promising solution, as evidenced by the surge of reported 2D heterostructures functioning as neuromorphic computing devices\cite{kim2020magneto, wang20202d}, as well as using skyrmions\cite{felser2022topology,yokouchi2022pattern,song2020skyrmion}.\,Another area is for quantum computing, concepts such as using 2D magnetic heterostructures with superconductors towards topological superconductivity\cite{palacio2019atomic,kezilebieke2020topological} and skyrmion qubits\cite{psaroudaki2021skyrmion,xia2023universal}, with scalibility, offer easy electric control and new functionality.\,In this review, we will first discuss the large family of existing 2D mangets, including ferromagnet, antiferromagnet, and multiferroics, then discuss the multifunctionality in 2D magnetic heterostructures. Based on these heterostructures, 2D devices for spintronics will be discussed, with a targeting for applications in high-performance and high-density magnetic memory, as well as neuromorphic and quantum computing.

\section{Library of 2D Magnets}

\begin{figure*}[ht]
\begin{centering}
 \includegraphics[width=0.85\textwidth]{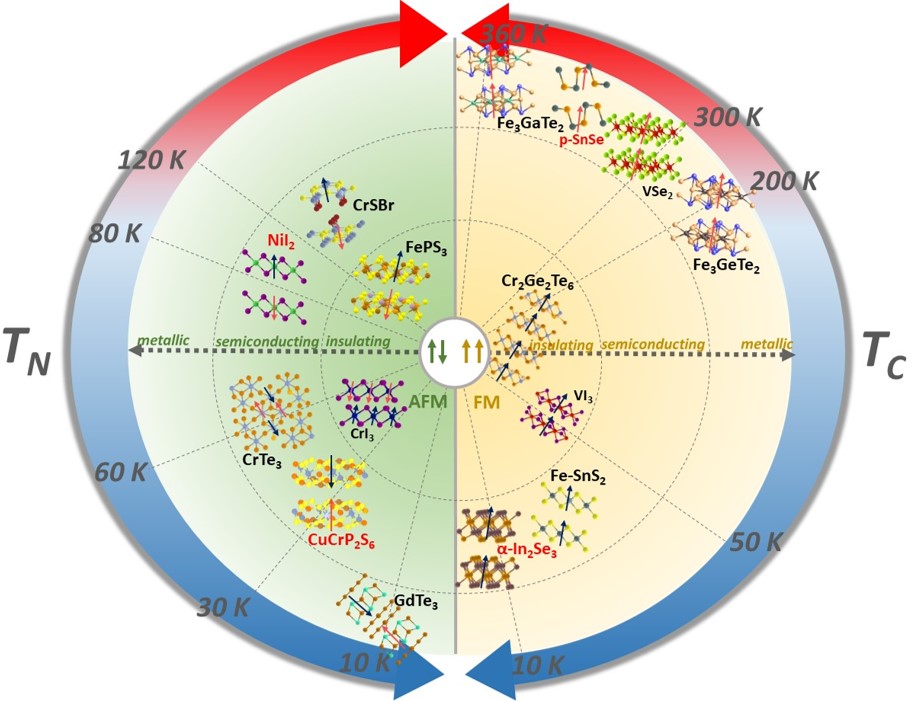} 
\par\end{centering}
\caption{Illustration of 2D magnetic materials, showing the critical temperature and electrical properties.\,Arrangement is in accordance with their magnetism, conductivity and critical temperature. Ferromagnetic and antiferromagnetic materials are characterized by Curie and N\'eel temperature, respectively. They are divided into three dashed spheres based on conductivity: smallest (insulating), medium (semiconducting), large (metallic). Multiferroics are highlighted by red color.}\label{Fig1} 
\end{figure*}

The hallmark of 2D magnetism is the existence of an ordered arrangement of magnetic moments over macroscopic length scales at any finite temperature, with a spontaneous breaking of time-reversal symmetry. Since first discovery in 2017, a rich collection of layered magnetic materials covering a wide spectrum of magnetic properties has been reported, as shown in Figure \ref{Fig1}. Many of these materials are semiconductors with bandgaps covering the near-infrared to the ultraviolet spectral range, whereas a few are metallic.

Ferromagnetic (FM) order exists in both layered transition metal halides and chalcogenides, when the exchange interactions favor both intralayer and interlayer parallel spin alignments. The Curie temperature $T_C$ is known as the temperature above which the material undergoes a magnetic transition from a ferromagnet to a paramagnet. 
While antiferromagnetic (AFM) materials have a rich variety of magnetic orders, they can be broadly divided into two types: those with intralayer FM order and interlayer AFM order, and others with intralayer AFM order. Antiferromagnets (antiferros) are magnetic ordered, yet with zero magnetization, making them immune to external field perturbations. Their AFM order is established by the exchange interactions between the spins, which produces an exchange field in the order of $10^2$ to $10^3$ T and give rise to a THz spin precession frequency. 
In the following, we discuss some magnetic materials under extensive studies, with physical properties promising for device design. 

Multilayer CrI$_3$ exhibits interlayer AFM behavior, while monolayers display FM ordering. The atomically thin CrI$_3$ flakes were successfully prepared by mechanical exfoliation, and the monolayer FM properties were subsequently demonstrated using a low-temperature microscopic magneto-optical Kerr effect (MOKE) measurement setup\cite{huang2017layer}.
The transition temperature $T_{C}$ for monolayer was measured at 45 K, slightly lower than the bulk $T_{C}$ of 61 K\cite{mcguire2015coupling,dillon1965magnetization}. The bilayer CrI$_3$ shows AFM behaviour due to a weak AFM interlayer coupling, which is different from monolayer. The interlayer magnetic coupling can be tuned by external field, either magnetic or electric fields. If an out-of-plane magnetic field is applied, with a field strength of $\sim$ 0.65 T, it can flip the magnetization in one layer and change the bilayer from AFM to FM state. Gate voltage applied through electric way can also help switch CrI$_3$ between FM and AFM states, with more energy-efficiency\cite{huang2018electrical}.

The intrinsic ferromagnetism in 2D Cr$_2$Ge$_2$Te$_6$ (CGT) layers was reported around the same time as CrI$_3$\cite{gong2017discovery}, and it is also revealed by MOKE microscopy under a small stabilizing magnetic field of 0.075 T. CGT is a FM material with a band gap around 0.8 eV\cite{hao2018atomic}. Its electrical property was characterized with a two-terminal resistivity of around M$\Omega$ at room temperature\cite{xing2017electric}, showing the electrically resistive nature. Thus few work can be done to study its transport properties. Additionally, CGT thin layers exhibit a thickness-dependent $T_C$, from a bilayer value of about 30 K to a bulk limit of 68 K. This thickness dependence of $T_C$ exists also in as-grown thin magnetic films, which shows more pronounced dependence\cite{zhang2019understanding}.  

These two materials mentioned above have relatively low $T_C$, with respect to room-temperature memory device.\,To market application of 2D magnets in spintronic devices, the  emphasis is on developing 2D materials with stable magnetic properties at room temperature. Among all 2D magnets, Fe-based materials usually have higher $T_C$, e.g. Fe$_3$GeTe$_2$ (FGeT). Its itinerant ferromagnetism was observed to persist down to the monolayer, due to its strong perpendicular magnetic anisotropy (PMA) with a uniaxial magnetocrystalline anisotropy constant $K_u$=1.46$\times 10^7$ erg/cm$^3$.\cite{leon2016magnetic} This anisotropy energy is about two or three orders of magnitude larger than CrI$_3$ ($K_u$=4.3$\times 10^4$ erg/cm$^3$ at 50 K)\cite{richter2018temperature} and CGT ($K_u$=1$\times10^5$ erg/cm$^3$)\cite{liu2019anisotropic} case. In bulk FGeT, the $T_C$ is $200-220$ K, depending on Fe deficiency. Although this value is diminished in atomically thin FGeT flakes, it can be significantly elevated to reach room temperature by employing liquid gating\cite{deng2018gate}, showing electrically tunable nature of these 2D magnetism.

Another example is VSe$_2$, with monolayer grown by molecular beam epitaxy (MBE) method.\,This material has been shown to maintain its FM ordering at 330 K\cite{bonilla2018strong}. Exceptionally, a recent discovered material Fe$_3$GaTe$_2$ (FGaT), with similar crystal structure to FGeT, exhibits as an above room temperature 2D metallic ferromagnet\cite{zhang2022above}. It has been demonstrated a record-high $T_{C}$ ($\sim$350-380 K) as well as a robust PMA. These outstanding magnetic properties render Fe$_3$GaTe$_2$ a widely favored material for practical magnetoelectronic and spintronic applications.


Beyond magnetic properties, these 2D magnets also provide a unique platform for the realization of topological magnetic materials, to study the interplay between magnetism and topology.\,MnBi$_2$Te$_4$ and its derivative compounds have received focused attention recently for their inherent magnetic order and the rich, robust and tunable topological phases\cite{wang2020topological}.\,MnBi$_2$Te$_4$ has an AFM interlayer coupling, while a FM order exists in each single layer, with out-of-plane easy magnetic axis. The interplay between the magnetic structure and the topological nontrivial bands endows the materials with rich topological phases, e.g. quantized anomalous Hall effect above 1.5 K\cite{liu2020robust}, which has higher temperature than magnetically doped BST samples\cite{pan2020observation}. In odd-layer flakes, quantum anomalous Hall effect was observed with quantized plateaus of Hall resistivity as well as diminishing longitudinal resistance\cite{deng2020quantum}, absence of any applied magnetic fields.\,This is extremely promising for low-power devices with dissipationless edge states.\,These materials and their heterostructures provide a neat, simple and versatile way to fuse topological bands with magnetic order, which are promising to help push quantum anamolous Hall effect towards the liquid nitrogen temperature. Possession of topology contributes to improving the performance of electronic components, by enabling high fault tolerance with topological protection, which is a key factor desired in quantum computing. 

Additionally, 2D multiferroic materials are actively studied for their cross-coupling effects. This class of materials exhibit multiple ferroic properties simultaneously, including ferromagnetism, ferroelectricity, and ferroelasticity. Multiferroics hold great promise for the development of innovative, multifunctional devices, with enhanced device performance and scaled down size in vdW case. The discovery of type-II multiferroic order in a single atomic layer of the transition-metal-based layered material NiI$_2$ was reported recently\cite{song2022evidence}, which exhibits an inversion-symmetry-breaking magnetic order and directly induces a ferroelectric (FE) polarization. Later, intriguing in-plane electrical and magnetic anisotropies in layered multiferroic CuCrP$_2$S$_6$\cite{wang2023electrical} has generated much excitement owing to coexsitence of antiferroelectric and antiferromagnetism, along with strong polarization-magnetization coupling. These multiferroics enable magnetization affected by electric field, and vice versa. This functionality has motivated fundamental and applied research in applying multiferroics for voltage controlled magnetic anisotropy, towards low-power switching devices. 

Overall, the current main thrusts for 2D magnets search are (1) room-temperature atomically thin magnet, for high-density and high-performance memory; (2) FM or AFM semiconductors with a tunable band gap and a stability, which would be an ideal platform to investigate the magneto-electric coupling between electrons and spin; (3) exotic topological properties, like searching for high-temperature quantum anomalous Hall insulators; (4) large-scale thin films for device integration. Although 2D exfoliated films offer high quality of layers, the limitation on the lateral size normally to hundreds of micrometers still prohibits their applications in the integration level. Efforts are working on to obtain high-quality wafer-scale 2D magnetic heterostructures.

\section{2D Magnetic Heterostructures and Their Interface} 

\begin{figure*}
\begin{centering}
 \includegraphics[width=0.8\textwidth]{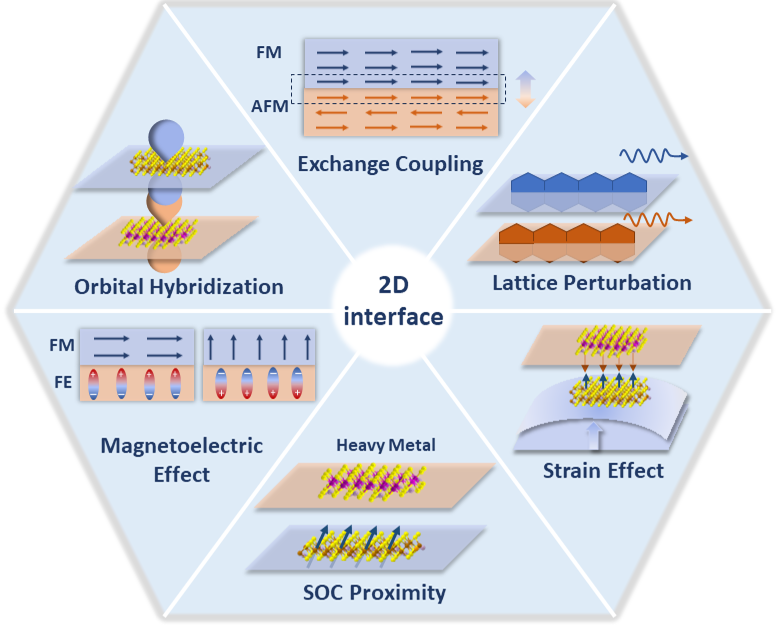} 
\par\end{centering}
\caption{Schematic mechanisms of  2D interface in magnetic heterostructures, including orbital hybridization, exchange coupling, lattice perturbation, strain effect, SOC proximity and magnetoelectric effect.}\label{Fig2} 
\end{figure*}

2D magnetic heterostructure is a material system composed of two or more distinct layers of atomically thin magnetic materials stacked on top of each other. These layers can have different magnetic properties or orientations. The use of 2D magnetic heterostructures is mainly for two purposes. One is to expand the functionality of these structures for both experimental investigations and future practical applications. The other is, the proximity effects within heterostructuers give rise to valuable phenomena that enhance the magnetic properties. When combining with magnetism, the interface has even more intriguing physics and phenomena as shown in Figure \ref{Fig2}, compared to the charge-based system.  

Firstly, the orbital hybridization at the interface can not only induce the exchange interaction at the interface, but also modify the electronic and magnetic properties in adjacent layers by impacting their band structure and orbital characters. Established orbital hybridization is a new degree of freedom for controlling interfacial Dzyaloshinskii-Moriya interaction (DMI), a short-range anti-symmetric exchange interaction\cite{zhu2022critical}.     
Taking CGT/FGeT heterostructure for examples, the topological Hall effect arising from magnetic skyrmions was observed, which are due to interfacial DMI\cite{wu2022van}. 

Another proximity effect is exchange coupling, which refers to the interaction between magnetic moment of different layers within the heterostructure.\,It's crucial for the applications in magnetic memory and spintronic devices, like spin valve, MTJ and magnetic sensors based on giant magnetoresistance (GMR) or tunneling magnetoresistance (TMR). As reported, large exchange bias observed in a FGeT/antiferro interface\cite{wu2022manipulating}, which is tunable by magnetic field cooling. Antiferro, with zero magnetization, shows robustness against external perturbations and THz fast spin procession. These makes coupling to antiferro materials extremely promising.   

Following is lattice perturbation, which distorts the crystal lattice structure at the interface, thus tuning the magnetic behaviours. Examples include applying electric voltages and vertical stacking to form the Moir\'e lattice\cite{yang2023moire}. This can lead to either the control of magnetization or the emergence of new exotic effects like Moir\'e skyrmions\cite{tong2018skyrmions}.

 Different from above mechanism, strain effect arises from mechanical deformation, leading to the change of electronic band structures. The methods to induce strain include lattice mismatch, utilization of flexible, patterned, or piezoelectric/FE substrates, as well as an atomic force microscopy tip or even the presence of bubbles at the interface during sample assembly. 

Spin-orbit coupling (SOC) refers to the interaction between the spin and orbital angular momentum of electrons across different layers within the heterostructure. This can play an important role when 2D magnet is interfaced with heavy metal element. An example of this is the emergence of skymions in WTe$_2$/FGeT heterostructure\cite{wu2020neel} through SOC coupling. This SOC is also crucial to realize spin-orbit torque (SOT) or spin-transfer torque (STT) in magnetic random-access memory (MRAM). 

The last but significant effect is the magnetoelectric effect, which occurs when a FM layer interfaces with a FE or multiferroic layer. The magnetism of the FM material can be switched by the electric polarization in the FE layer, and vice versa. FE control of magnetism is particularly interesting due to its energy-efficient and nonvolatile nature. One example is to use a thin FE polymer to open and close the CGT magnetic loop\cite{liang2023small}. \begin{table}[h]
\begin{center}
\begin{tabular}{|p{2cm}|c|c|c|c|} \hline
\multicolumn{2}{|c|}{Material} & Thickness (nm) & Polarization ($\mu$C/cm$^2$) & Coercive field (MV/cm) \\ \hline
\multirow{3}{*}{2D FEs} & CuInP$_2$S$_6$ & N/A & 2.55 & 0.077 \\
& $\alpha$-In$_2$Se$_3$ & 5 & 2-11 & 0.2 \\ 
& h-BN & N/A & 0.09 & 2 \\\hline
\multirow{4}{*}{Other FEs} & Hf$_x$Zr$_{1-x}$O$_2$ & 0.5-30 & 5-50 & 1-3 \\
& Al$_{1-x}Sc_xN$ & <10 & 100-150 & 3-7\\
& Al$_{1-x}B_xN$ & 500 & >125 & 5-6 \\
& Zn$_{1-x}Mg_xO$ & 500 & 25-100 & 2-3 \\ \hline
\end{tabular} 
\caption{Comparing FE properties between different classes of 2D and thin materials.\label{tab1}}
\end{center}
\end{table}
An alternative promising approach to generate a substantial magnetoelectric effect involves the stacking 2D FM layer with recently emerging fluorite\cite{1hfO,2ZrHfO,3FEFET} and wurtzite FE\cite{4SCFE,5FEboron,6FEmagnetism} materials. These CMOS-compatible FE materials, such as hafnium-dioxide, zinc-magnesium-oxide, or aluminum-scandium-nitride, offer a unique advantage over conventional perovskites due to their remarkable scalability, while providing very large polarization and coercive fields\cite{7FEwurttizitefluorite}. Moreover, fluoride FEs with sub-nm thickness\cite{8FEbinaryoxide,9HfO2} and wurtzite FEs with sub-10 nm thickness\cite{10FEnitirde} have been demonstrated, and their integration with 2D materials is explored to create novel memory devices are explored\cite{11nitride}. In comparison to 2D FEs\cite{12recentprogress,13vdwFE} as shown in Table \ref{tab1}, these 3D yet atomically thin materials offer one to two orders of magnitude higher polarization, significantly amplifying magnetoelectric effects at the interface with FMs\cite{14NiHfO}. Additionally, they boast coercive fields two to three orders of magnitude higher compared to 2D FE counterparts, enabling the attainment of higher switching voltages and improved interstate isolation,  which is desirable for memory applications.

The high-quality and seamless 2D interfaces offer a platform for introducing new functionalities into 2D devices, thereby enabling the exploration of novel physics and effects. Of particular interest is the pursuit of ultra-fast, ultra-compact, and energy-efficient spintronics through precise interface engineering. By leveraging the interface, it becomes possible to amalgamate THz spin procession in AFM materials, nonvolatile control via 2D FEs, and scalable configurations into a single device. This development holds significant promise for advancing the field of spintronics in the future.

\section{2D Spintronics}
\subsection{Device for Magnetic Memory}

\begin{figure*}
\begin{centering}
 \includegraphics[width=1\textwidth]{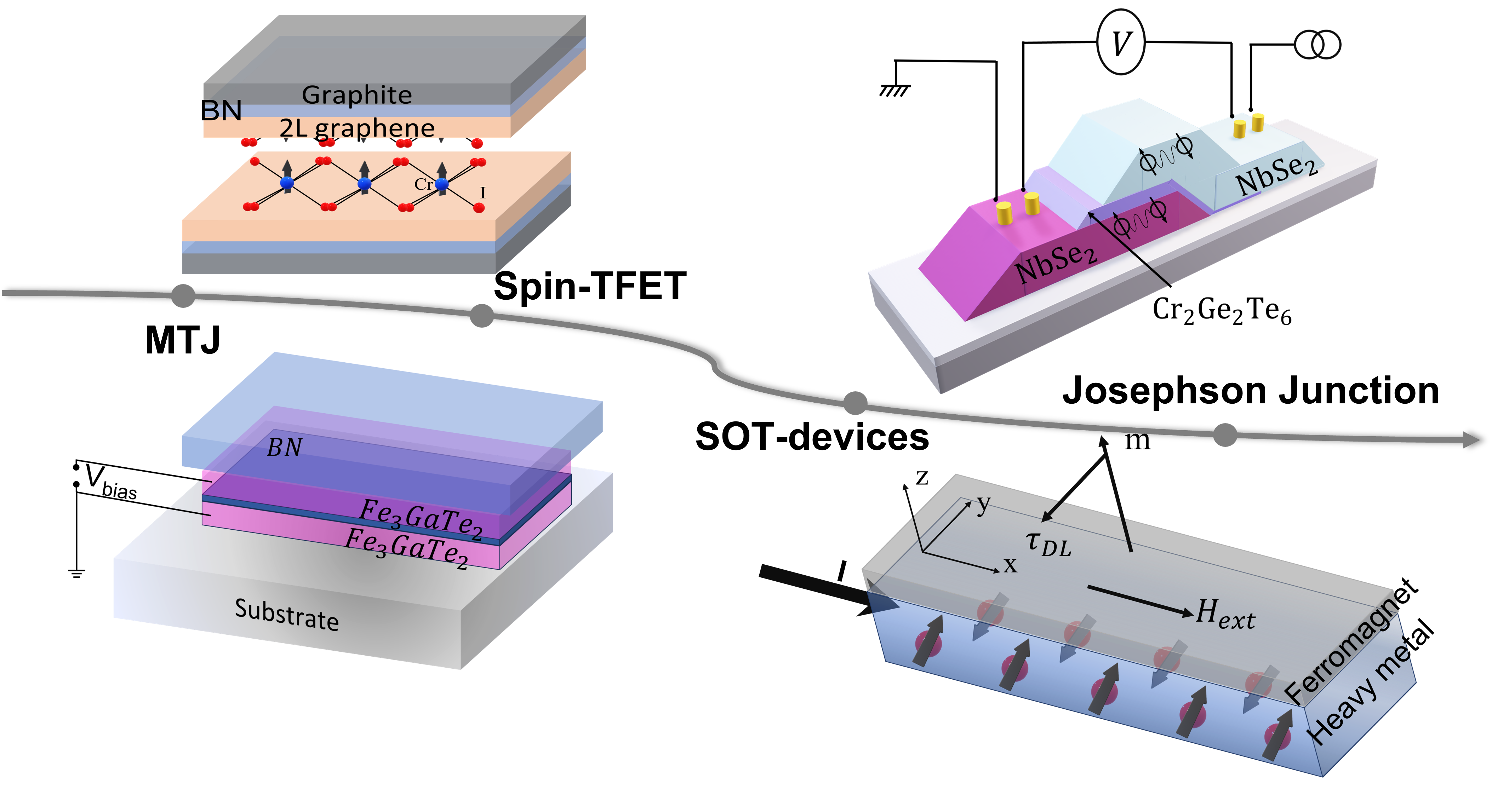} 
\par\end{centering}
\caption{Schematic of 2D magnetic heterostructures for magnetic memory. Examples include magnetic tunneling junction (MTJ), spin-tunnel field-effect transistor (spin-TFET), spin-orbit torque (SOT) devices and Josephson junction.}\label{Fig3} 
\end{figure*}

Introducing magnetic elements into memory devices provides substantial advantages, particularly in achieving non-volatile memory functionality. The utilization of 2D magnetic heterostructures, comprised of multiple layers of 2D materials with unique properties, takes these advantages even further. It enhances energy efficiency, boosts data storage density, and enriches the range of functionalities achievable within a single memory device, as shown in Figure \ref{Fig3}.

MTJ is a fundamental component in modern memory devices. It consists of two FM layers separated by a thin insulating barrier. The core principle governing MTJ operation is the TMR effect, wherein the electrical resistance of the junction changes based on the relative alignment of magnetization in the two FM layers. In contrast to traditional bulk ferros such as Fe$_3$O$_4$ and Co, employing atomically thin layered magnets for MTJs offers the potential to significantly enhance their TMR, since TMR heavily depends on interface quality.\,Additionally, this approach relaxes the stringent lattice-matching requirements associated with epitaxy growth and enables high-quality integration of dissimilar materials with atomically sharp interfaces. 

Among the existing 2D magnets, CrI$_3$ stands out due to its layered AFM ordering and relatively electrically insulating properties, making it a desirable choice as tunnel barrier in MTJs. One spin-filter MTJ design involves a heterostructure consisting of h-BN encapsulated graphene/CrI$_3$/graphene layers\cite{song2018giant}. This device demonstrated remarkable TMR values of 530\%, 3200\%, and 19,000\% for bilayer, trilayer, and four-layer CrI$_3$ structures, respectively. The highest value of the four-layer structure greatly exceeds typically achieved in conventional bulk material MTJs\cite{ikeda2008tunnel,parkin2004giant}.\,While MTJs based on CrI$_3$ exhibit promising properties in TMR, it's crucial to acknowledge their operational constraints - cryogenic low temperature for operation and need of magnetic fields. In response, ongoing research focuses on advancing MTJs by employing materials suitable for room-temperature operation. A notable candidate is FGaT, which boasts a high $T_C$ of 380 K, as well as operation temperature above room temperature. This characteristic overcomes the limitation, greatly facilitating the realization of practical spintronic devices.\,A room-temperature MTJ based on a FGaT/WSe$_2$/FGaT heterostructure was successfully demonstrated\cite{zhu2022large}, showing significant TMR of 85\% at room temperature.
 
Spin field-effect transistor (Spin-FET) is a new type of spintronics device towards non-volatile memory. This device utilizes the spin, rather than charge, in the flow of electrical current, which could potentially offer non-volatile data storage and improved performance compared with traditional FETs. The application of a magnetic tunnel barrier in MTJ can be extended to Spin-FETs, leading to a noval device - the spin tunnel FET, e.g. dual-gated graphene/CrI$_3$/graphene tunnel junctions\cite{jiang2019spin}, with similar structure as spin-filter MTJs. This device achieves an impressive high-to-low conductance ratio of nearly 400\% by electrical modulation of magnetization configurations. However, it shares a common constraint with CrI$_3$-based MTJs, requiring low operational temperatures and magnetic fields. Despite these challenges, the search for high-temperature semiconducting magnets holds the key to advancing the development of this device, potentially overcoming existing limitations and paving the way for future applications.

Utilizing SOC, SOT devices are engineered to manipulate the magnetization of a magnetic layer by an spin-polarized current. Compared to traditional STT devices, SOT devices are anticipated to offer superior performance like easy stacking, operation in sub-ns writing and reading speed. Typically, a SOT device comprises two key components: a FM layer and a non-magnetic heavy metal (HM) layer. SOT efficiency heavily relies on the interface quality. Thus, 2D magnets can be promising building blocks for SOT devices, like CGT, CrI$_3$, and FGeT with intrinsic PMA. Even in monolayer form, these materials maintain their PMA, making them highly responsive to external stimuli. 
In the initial designs of 2D magnetic SOT devices, structures like Pt/FGeT\cite{alghamdi2019highly} and Ta/CGT\cite{ostwal2020efficient} were featured, showing significant improvements in SOT switching efficiency with the incorporation of 2D FM materials. Creating all-vdW heterostructures can further enhance the switching efficiency. For instance, a FGeT/WSe$_2$ heterostructure was realized\cite{shin2022spin}, achieving a remarkably low switching current density of 3.90$\times10^6$ A/cm$^2$ at 150 K,  which is comparable or even better to that of conventional thin films system\cite{xie2022magnetization,kang2021critical}.
Recent research efforts have been directed towards achieving reliable vdW SOT operation at room temperature. One work\cite{li2023room} incorporates high $T_C$ FGaT to achieve room temperature operation, with a FGaT/Pt bilayer structure. Currently, this approach still needs a high switching current density of 1.3$\times$10$^7$ A/cm$^2$. Another notable approach involves a wafer-scale FGeT/Bi$_2$Se$_3$ layered heterostructure\cite{wang2023room}, where the topological insulator Bi$_2$Se$_3$ plays a crucial role in raising the \textit{T$_C$} of FGeT to room temperature through exchange coupling. This approach has yielded promising results, including a high damping-like SOT efficiency of $\sim$2.69 and a notably low switching current of 2.2$\times$ 10$^6$ A/cm$^2$ at room temperature. This development marks a significant step toward practical SOT devices for real applications.

Apart from conventional memory, Josephson junction (JJ) combines the quantum phenomena with memory device technology, opening the door to the development of superconducting Random Access Memories (RAMs)\cite{semenov2019very}. JJs are typically composed of two superconductors (SCs) separated by a thin insulating or non-superconducting barrier. JJ based on 2D ferro/SC heterostructure presents unique opportunity to explore the interaction between superconductivity and ferromagnetism.\,For instance, one approach involves constructing a JJ by introducing a few-layer FM insulator, CGT, between two layers of SCs NbSe$_2$\cite{ai2021van}. This JJ demonstrates a hysteresis behavior in the critical current, which is induced by the remanence of the magnetic barrier. Following this, another lateral JJ emerged, composed of a FGeT, laterally interconnected between two layered spin-singlet SCs NbSe$_2$\cite{hu2023long}. This SC/ferro/SC heterostructure successfully sustains skin Josephson supercurrents with a remarkable long-range reach over 300 nanometers (nms).

\subsection{2D Magnetic Heterostructures for Neuromorphic Computing}           

\begin{figure*}
\begin{centering}
 \includegraphics[width=0.9\textwidth]{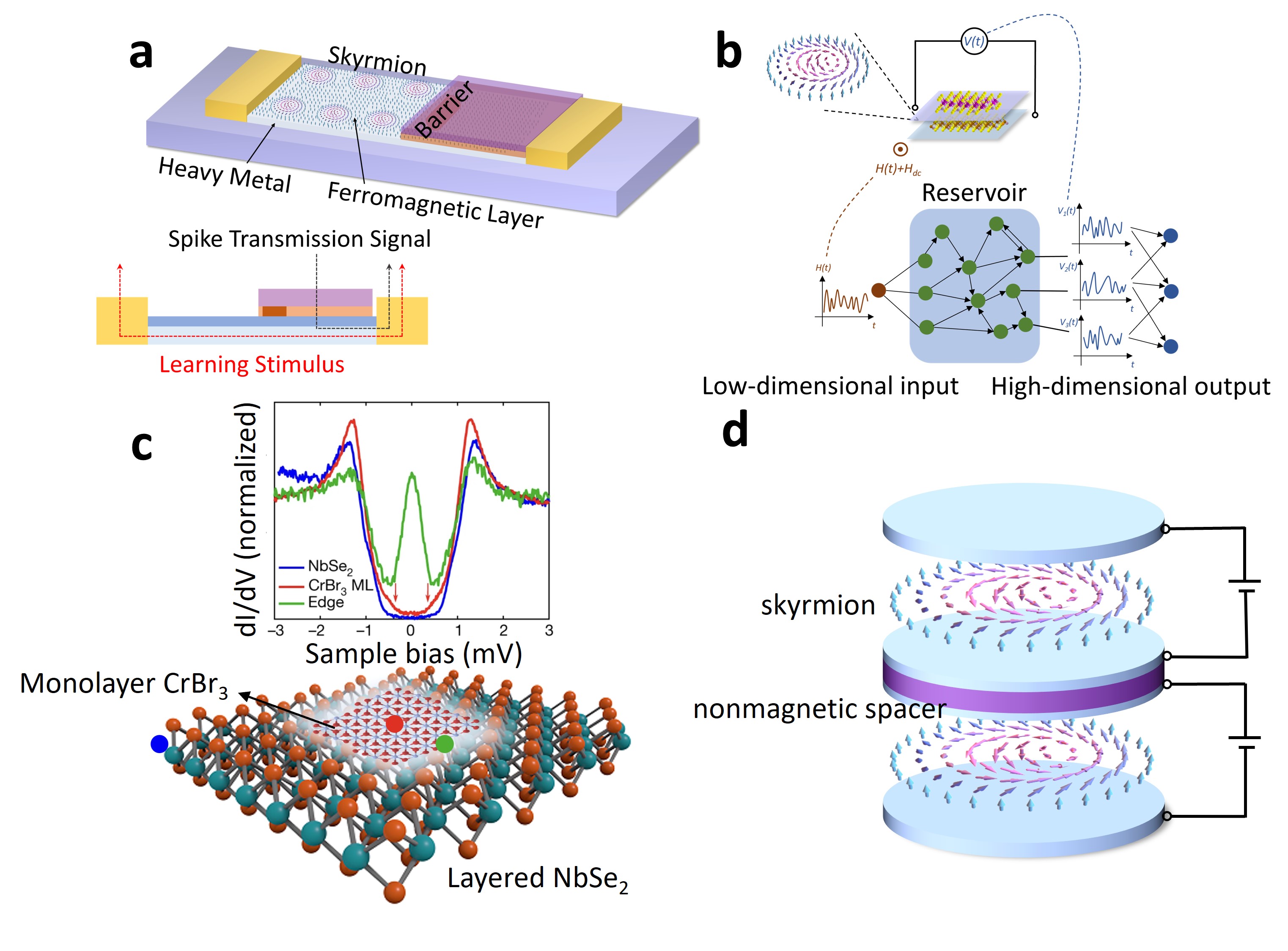} 
\par\end{centering}
\caption{ (a) Schematic of the skyrmionic synaptic device. Skyrmions generated within 2D ferro/heaavy metal heterostructure. Bidirectional stimulus applied on heavy metal layer inducing the spin current in FM layer. (b) Schematic of skyrmion neuromorphic computer based on reservoir computing. Low-dimensional voltage input applied to 2D skyrmion systems yields high-dimensional magnetic output\cite{yokouchi2022pattern}.\,(c) Schematic of CrBr$_3$/NbSe$_2$ topological superconductor.\,$dI/dV-V_\textrm{bias}$ spectrum charaterized on the NbSe$_2$ substrate (blue), the middle of CrBr$_3$ island (red) and the edge of CrBr$_3$ island (green)\cite{kezilebieke2020topological}.\,(d) Schematic of skyrmion qubit concept based on magnetic heterostructure. 
}\label{Fig4} 
\end{figure*}  

Neuromorphic computing is an emerging field of artificial intelligence that seeks to design computer systems inspired by the human brain's neural architecture. 
Owing to the atomic thickness, dangling-bond-free surfaces and high mechanical robustness, 2D materials and heterostructures are extensively investigated for neuromorphic computing devices, showing great promise for high-performance artificial neurons and synapses\cite{tong20212d,huh2020memristors,wang2022review,yan2022progress,tian2021recent}.\,The integration of magnetic elements within 2D materials to form 2D magnetic heterostructures takes this field a step further. These magnetic elements provide non-volatility, longevity, and high-density information storage capabilities. Furthermore, 2D magnetic heterostructures have the capability to generate skyrmions\cite{wu2022van,wu2020neel}, which are topological spin structures with unique topology towards more compact, energy-efficient, fault-tolerant and resilient neuromorphic devices\cite{song2020skyrmion,yokouchi2022pattern,yu2020voltage}.
Figure \ref{Fig4}a schematics an artificial synapse device concept based on skyrmions\cite{huang2017magnetic}. The primary component of the device is a 2D ferro/heavy metal heterostructrue, which forms a nanotrack for skyrmions motion. The system replicating the operations of biological neurons and synapses through the following processes: in spike transmission mode, the pre-neuron's spike is modulated by the synaptic device's weight (reading through magnetoresistance), leading to the generation of a post-synaptic spike current.\,During the learning phase, a bidirectional charge current through the heavy metal layer injects a vertical spin current in FM layer, driving the skyrmions into (or out of)  the postsynaptic region. This dynamic process effectively adjusts the synaptic weight, closely mirroring the potentiation and depression mechanisms observed in biological synapses. Additionally, the resolution of the synaptic weight can be adjusted based on the nanotrack width and the skyrmion size, which enriches the flexibility and tunibility of the skyrmion-based synapse. Another example for the application of 2D skyrmions is the reservoir computing\cite{yokouchi2022pattern}, a framework derived from recurrent neural network theory. The essence of reservoir computing is the nonlinear transformation of input into high-dimensional outputs.\,A physical system capable of implement the reservoir part should possess both complex nonlinearity and memory effect (or equally hysteresis) while also exhibiting short-term properties\cite{du2017reservoir,nakane2021spin,milano2022materia}. Thus, magnetic skyrmions emerge as a promising candidate. In Figure \ref{Fig4}b, 2D skyrmion system performs as the "resvervoir part", nonlinearly converting the one-dimensional time-series input $H_\textrm{AC}(t)$ to the linearly independent $N$-dimensional time-series outputs $V(t)$. This design provides a guideline for developing energy-saving and high-performance skyrmion neuromorphic computing devices.

\section{2D Heterostructures for Quantum Computing}

2D magnetic heterostructures hold promise for quantum computing due to their unique electronic properties, controllability, and potential for creating novel quantum states. These interfaces can give rise to emergent quantum phenomena, broadening the horizons for quantum information processing.\,Notably, topological quantum computing benefits significantly from the topological protection and qubit safeguarding that 2D magnet/magnetic heterostructure can offer.
An illustrative example is the observation of 2D topological superconductivity within vdW heterostructures that combine the ferro CrBr$_3$ with the SC NbSe$_2$\cite{kezilebieke2020topological}. As Figure \ref{Fig4}c shows, the monolayer CrBr$_3$ island was grown on the NbSe$_2$ layer. The $dI/dV$ spectrum was characterized at different sites of this structure: NbSe$_2$ substrate (blue), the middle of the CrBr$_3$ island (red) and the edge of the CrBr$_3$ (green). Among them, a peak of conductance localized at $E_F$ can be clearly seen for the edge of magnetic island, which exhibits as a hallmark of topological superconductivity for this structure.
Furthermore, owing to the 2D layered structure, these edge modes can be readily accessed and manipulated using various external stimuli, including electrical, mechanical, and optical methods. This feature enhances the promise of these heterostructures for integration into components of topological quantum computing systems.

Another application of 2D magnetic heterostructures in quantum computing is the implementation of skyrmion qubit. 
Unlike other proposed qubit systems such as trapped atoms, quantum dots, and photons, magnetic skyrmions offer a topologically protected structure that shows promise in resisting external perturbations and faults.\,Moreover, skyrmion qubits hold additional potential in addressing challenges related to control and scalability, further enhancing their viability in quantum information processing. In Figure \ref{Fig4}d, a skyrmion qubit concept is illustrated, utilizing a bilayer magnetic materials platform\cite{psaroudaki2021skyrmion}. In this setup, quantum information is stored within the quantum degree of helicity, and the logical states can be dynamically adjusted through the manipulation of electric and magnetic fields.

\section{Conclusions and perspectives}
For decades, 2D magnetism has been a captivating subject, especially in the context of the Mermin-Wagner-Hohenberg (MWH) theorem, which predicts that thermal fluctuations will disrupt long-range magnetic order in 2D systems at any finite temperature, following the isotropic Heisenberg model. Extensive theoretical and experimental studies have been undertaken to unravel causes of long-range ordering in 2D systems. Beyond magnetization, other physical properties, including polarization, that may exhibit potential long-range order within 2D systems hold the promise of pushing the limits of Moore's law.\,The inherent atomic-layer cleavability and magnetic anisotropy of 2D magnets may help mitigate spin fluctuations in the face of short-range interactions, thus paving the way for the emergence of 2D magnetism.\,While the duality in electromagnetism remains a profound aspect of fundamental physics, the recent discovery of 2D multiferroics further reinforces this concept, as ferroelectricity is analogous to ferromagnetism. These discoveries pose thought-provoking questions, such as the existence of an electric counterpart to the MWH theorem.\,Considering dipole-dipole interactions, the absence of polarization becomes a relevant consideration. Much like magnetic skyrmions are topological whirls in magnetization, one might comtemplate the existence of a FE equivalent.

The allure of 2D magnets extends beyond fundamental physics and into the realm of novel device fabrication. Significant progress has been made in comprehending their fundamental properties, and the demonstration of devices is in its nascent stages.\,We have witnessed the expansion of the materials library and glimpsed the potential device functionalities. However, achieving integration of these materials into functional devices requires large-area wafer-scale materials growth. Encouraging strides have been made through techniques such as MBE and atomic layer deposition (ALD), e.g. large-scale monolayer CrBr$_3$ with MgO as a passivation layer from ALD \cite{galbiati2020very} and wafer-scale FGeT on Bi$_2$Te$_3$ from MBE\cite{wang20202d}. Challenges still persist at the device level, particularly in advancing dielectric and contact interfaces. The self-passivated nature of monolayer 2D magnets will necessitate seeding for the deposition of dielectrics through methods like ALD. It may lead to non-ideal interfaces, limiting device performance compared to the best laboratory data that employs crystalline 2D insulators like h-BN. Similar challenges are encountered with electrical contacts, as they only partially conform to industry specifications and have not yet reached the level of readiness required for manufacturing. Addressing these crucial manufacturing bottlenecks will pave the way for a significant enhancement in chip functionality, heralding a new era of 2D magnet applications characterized by increased device complexity.

Moreover, in the context of neuromorphic computing, a fundamental challenge lies in improving the endurance of resistance switching. Achieving material uniformity is essential to create massively connected device arrays capable of mimicking the hyper-connectivity and efficiency of the brain. Computational methods will be employed to guide experimental studies and optimize memristive devices for maximum performance. 

For using 2D magnetic heterostructures in topological quantum computing, quantum states can benefit from topological protection, safeguarding qubits against external perturbations and errors, in contrast to standard quantum computing. Efforts in topological superconductivity and skyrmion qubits present a unique opportunity to explore the interplay between topology, magnetism, and electronic properties, potentially yielding new discoveries and insights in physics, as well as paving the way for future quantum computing devices.

\bigskip
\textbf{References}
\bibliographystyle{apsrev}
\bibliography{review.bib}

\end{document}